%% file: main.tex
\documentclass[conference,10pt]{IEEEtran}
         
\usepackage{amssymb}
\usepackage{amsmath}
\usepackage{graphicx}
\usepackage[ruled]{algorithm2e}
\usepackage{bm}
\usepackage{romannum}
\usepackage{xcolor}

\begin{document}
\title{Learning How to Demodulate from Few Pilots \\ via Meta-Learning}
\author{\IEEEauthorblockN{Sangwoo Park}
\IEEEauthorblockA{\textit{School of Electrical Engineering} \\
\textit{KAIST}\\
Daejeon, South Korea}
\and
\IEEEauthorblockN{Hyeryung Jang, Osvaldo Simeone}
\IEEEauthorblockA{\textit{Dept. of Informatics} \\
\textit{King's College London}\\
London, United Kingdom }
\and
\IEEEauthorblockN{Joonhyuk Kang}
\IEEEauthorblockA{\textit{School of Electrical Engineering} \\
\textit{KAIST}\\
Daejeon, South Korea}
}

\maketitle

\begin{abstract}
Consider an Internet-of-Things (IoT) scenario in which devices transmit sporadically using short packets with few pilot symbols. Each device transmits over a fading channel and is characterized by an amplifier with a unique non-linear transfer function. The number of pilots is generally insufficient to obtain an accurate estimate of the end-to-end channel, which includes the effects of fading and of the amplifier's distortion. This paper proposes to tackle this problem using meta-learning. Accordingly, pilots from previous IoT transmissions are used as meta-training in order to learn a demodulator that is able to quickly adapt to new end-to-end channel conditions from few pilots. Numerical results validate the advantages of the approach as compared to training schemes that either do not leverage prior transmissions or apply a standard learning algorithm on previously received data.
\end{abstract}


\vspace{0.2cm}
\input{intro}  
\vspace{0.2cm}
\input{model}

\vspace{0.2cm}
\input{meta_training}

\vspace{0.2cm}
\input{experiments}  

\vspace{0.2cm}
\section*{Acknowledgment}
The work of S. Park and J. Kang was supported by the National Research Foundation of Korea (NRF) grant funded by the Korea government (MSIT) (No. 2017R1A2B2012698). The work of H. Jang and O. Simeone was supported by the European Research Council (ERC) under the European Union's Horizon 2020 research and innovation programme (grant agreement No. 725731).

\vspace{0.2cm}
\bibliographystyle{IEEEtran}






\end{document}

%% file: intro.tex
\section{Introduction}
\label{sec:intro}

For channels with an unknown model or an unavailable optimal receiver of manageable complexity, the problem of demodulation and decoding can potentially benefit from a data-driven approach based on machine learning (see, e.g, \cite{ibnkahla2000applications}, \cite{simeone2018very}). Demodulation and decoding can in fact be interpreted as classification tasks, whereby the input is given by the received baseband signals and the output by the actual transmitted symbols, for demodulation, or by the transmitted binary messages, for decoding. Pilot signals can hence be used as training data to carry out supervised learning of a given parametric model, such as  Support Vector Machines or neural networks. The performance of the trained ``machine" as a demodulator or a decoder generally depends on how representative the training data is for the channel conditions encountered during test time. 

The design of optimal demodulators and decoders is well-understood for many standard channels, such as additive Gaussian noise channels and fading channels with receive Channel State Information (CSI), making a data-driven approach undesirable. In fact, most communication links use pilot sequences to estimate CSI, which is then plugged into the optimal receiver with ideal receive CSI (see, e.g., \cite{ostman2018short}). However, this standard approach based on domain knowledge is inapplicable if: (\emph{i}) an accurate channel model is unavailable; and/or if (\emph{ii}) the optimal receiver for the given transmission scheme and channel is of prohibitive complexity or unknown. Examples of both scenarios are reviewed in \cite{ibnkahla2000applications}, \cite{simeone2018very} and include new communication set-ups, such as molecular channels, which lack well-established models, and links with strong non-linearities, such as satellite links with non-linear transceivers, whose optimal demodulators can be highly complex \cite{ibnkahla2000applications}, \cite{bouchired1998equalisation}. This observation has motivated a long line of work on the application of machine learning methods to the design of demodulators or decoders, from the 90s \cite{ibnkahla2000applications} to many recent contributions including \cite{o2017introduction} and references therein.

\begin{figure}
    \centering
    \includegraphics[width=0.85\columnwidth]{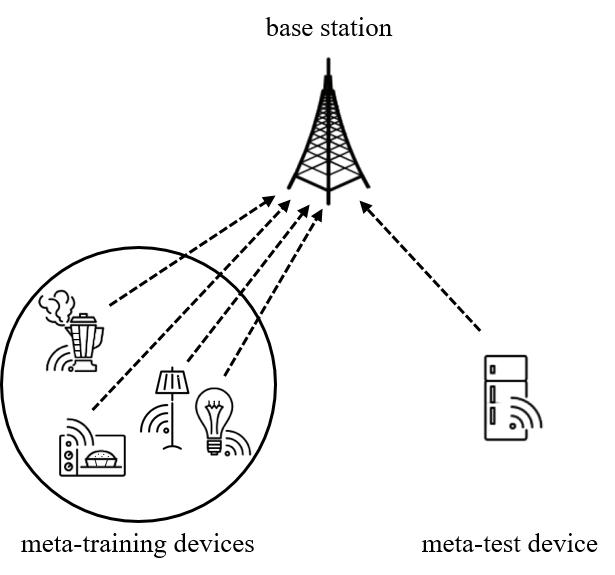}
    \caption{Illustration of few-pilot training for an IoT system via meta-learning.}
    \label{fig:iot_system}
\end{figure}

To the best of our knowledge, all the prior works reviewed above assume that training is carried out using pilot signals from the same transmitter whose data is to be demodulated or decoded. This generally requires the transmission of long pilot sequences for training. In this paper, we consider an Internet-of-Things (IoT)-like scenario, illustrated in Fig.~\ref{fig:iot_system}, in which devices transmit sporadically using short packets with few pilot symbols. The number of pilots is generally insufficient to obtain an accurate estimate of the end-to-end channel, which generally includes the effects of fading and of the amplifier's distortion \cite{helmy2017robustness}. We propose to tackle this problem using \emph{meta-learning} \cite{thrun1998lifelong}. 

Meta-training, also sometimes referred to as ``learning to learn", aims at leveraging training and test data from different, but related, tasks for the purpose of acquiring an inductive bias that is suitable for the entire class of tasks of interest \cite{thrun1998lifelong}. The inductive bias can take the form of a learning procedure or a prior over the model parameters \cite{grant2018recasting}. An important application is the acquisition of a learning algorithm or of a model prior that allows a quick adaptation to a new, related, task using few training examples, also known as \emph{few-shot learning} \cite{vinyals2016matching}. For instance, one may have training and test labelled images for binary classifiers of different types of objects, such as cats vs dogs or birds vs bikes, which can be used as meta-training data to quickly learn a new binary classifier, say for handwritten digits, from a few training examples. 

As illustrated in Fig.~\ref{fig:meta_4pam}, the key idea of this paper is to use pilots from previous transmissions of other IoT devices as meta-training in order to learn a demodulator that is able to quickly adapt to new end-to-end channel conditions from few pilots. This paper provides a solution based on the model-agnostic meta-learning (MAML) algorithm proposed in \cite{finn2017model} and presents numerical experiments for basic demodulation tasks.

%% file: model.tex
\section{Model and Problem}
\label{sec:model}

\subsection{System Model}

In this paper, we consider the IoT system illustrated in Fig.~\ref{fig:iot_system}, which consists of a number of devices and a base station (BS). For each device $k$, we denote by $s_k \in \mathcal{S}$ and $y_k$ the complex symbol transmitted by the device and the corresponding received signal at the BS, respectively. We also denote by $\mathcal{S}$ the set of all constellation symbols, as determined by the modulation scheme. The end-to-end channel for a device $k$ is defined as 
\begin{align} \label{eq:e2e-channel}
y_k = h_k x_k + z_k,
\end{align}
where $h_k$ is the complex channel gain from device $k$ to the BS, which we assume constant over the transmission of interest; $z_k \sim \mathcal{C}\mathcal{N}(0,N_o)$ is additive white complex Gaussian noise; and $x_k = g_k(s_k)$ is the output of a non-linearity accounting for the amplifier's characteristics of the IoT device \cite{helmy2017robustness}. As an example, a standard model for the non-linear function $g_k(\cdot)$ that assumes only amplitude distortion is \cite{bouchired1998equalisation}
\begin{align} \label{eq:non-linearity}
x_k = g_k(s_k) = \frac{\alpha_k |s_k|}{1 + \beta_k |s_k|^2} \exp(j \measuredangle s_k),
\end{align}
where $\measuredangle s_k$ represents the phase of symbol $s_k$, and $\alpha_k$ and $\beta_k$ are constants depending on the characteristics of the device. 

As detailed below, based on the reception of a few pilots from a device, we aim at determining a probabilistic demodulator that recovers the transmitted symbol $s$ from the received signal $y$ with high probability. The demodulator is defined by a conditional probability distribution $p_\theta(s | y)$, which depends on a trainable parameter vector $\theta$.

\subsection{Meta-Learning Problem} 

\begin{figure}[t!]
    \centering
    \includegraphics[width=0.7\columnwidth]{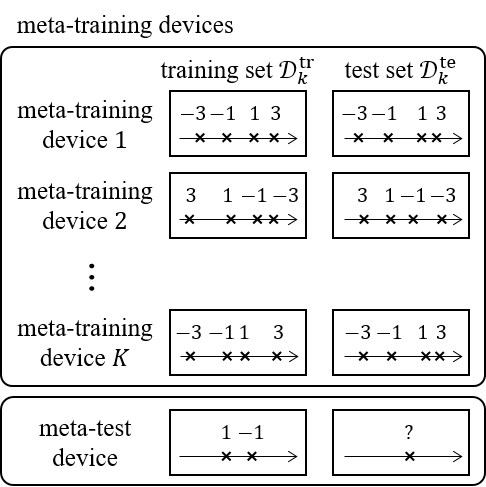}
    \caption{Illustration of meta-training and meta-test data for $4$-PAM transmission from set $\mathcal{S} = \left\{-3,-1,1,3\right\}$. The figure assumes $N=8$ pilot symbols divided into $N^\text{tr}=4$ for meta-training and $N^\text{te}=4$ for meta-testing, and $P=2$ pilots for the meta-test device. Crosses represent received signals $y_k^{(n)}$, and the number above each cross represents the corresponding label, i.e., the pilot symbol $s_k^{(n)}$.}
    \label{fig:meta_4pam}
\end{figure}

As mentioned, we are interested in learning an effective demodulator $p_\theta(s|y)$ for a given IoT device based on the transmission of a few pilot symbols. Following the nomenclature of meta-learning \cite{finn2017model}, we refer to the target device as the {\em meta-test device}. To enable few-pilot learning, the BS can use the signals received from the previous pilot transmissions of $K$ other IoT devices, which are referred to as {\em meta-training devices} and their data as \emph{meta-training data}. Specifically, as illustrated in Fig.~\ref{fig:meta_4pam}, the BS has available $N$ pairs of pilot $s_k$ and received signal $y_k$ for each meta-training device $k=1,\ldots,K$. The meta-training dataset is denoted as $\mathcal{D} = \{\mathcal{D}_k \}_{k=1,\ldots,K}$, where $\mathcal{D}_k = \{ (s_k^{(n)}, y_k^{(n)}): n=1,\ldots,N \}$, and $(s_k^{(n)}, y_k^{(n)})$ are the pilot-received signal pairs for the $k$th meta-training device. 

For the target, or the meta-test, device, the BS receives $P$ pilot symbols. We collect the $P$ pilots received from the target device in set $\mathcal{D}_\text{T} = \{ (s^{(n)},y^{(n)}): n=1,\ldots,P\}$. 
The demodulator can be trained using meta-training data $\mathcal{D}$ and the pilot symbols $\mathcal{D}_\text{T}$ from the meta-test device. 

Training requires the selection of a parametric model $p_\theta(s|y)$ for the demodulator. The choice of the parametric model $p_\theta(s|y)$ should account for the standard trade-off between expressivity of the model and overfitting \cite{bishop2006pattern, simeone2018brief}. A classical choice is a multi-layer neural network with $L$ layers, with a softmax non-linearity in the final, $L$th, layer. This can be written as
\begin{align} \label{eq:softmax-demod}
p_\theta(s|y) = \frac{\exp \Big( w_s^{(L) \top} f_{W^{(L-1)}} ( f_{W^{(L-2)}} (\cdot \cdot f_{W^{(1)}}(y) ) ) \Big) }{ \sum\limits_{s' \in \mathcal{S}} \exp \Big( w_{s'}^{(L) \top} f_{W^{(L-1)}} ( f_{W^{(L-2)}} (\cdot \cdot f_{W^{(1)}}(y) ) ) \Big) },
\end{align}
where $f_{W^{(l)}}(x) = \sigma (W^{(l)}x)$ represents the non-linear activation function of the $l$th layer with weight matrix $W^{(l)}$ of appropriate size, and $\theta = \{ \{W^{(l)}\}_{l=1,\ldots,L-1}, \{w_s^{(L)} \}_{s \in \mathcal{S}} \}$ is the vector of parameters. The non-linear function $\sigma(\cdot)$ can be, e.g., a ReLU or a sigmoid function. The input $y$ can be represented as a two-dimensional vector comprising real and imaginary parts of the received signal in \eqref{eq:e2e-channel}.

%% file: meta_training.tex
\section{Meta-learning Algorithm} 
\label{sec:meta}

In this section, we propose a meta-learning algorithm for the design of a demodulator given $P$ pilots of the target device. The approach uses the meta-training data $\mathcal{D}$ to identify a parameter vector $\theta$ that enables rapid adaptation based on the few pilots of the meta-test device. Note that this identification can be carried out in offline manner prior to the reception from the target device. The initial weight can be reused for multiple target devices.

As discussed in Sec.~\ref{sec:intro}, we view demodulation as a classification task (see, e.g., \cite{ibnkahla2000applications, simeone2018very}). Accordingly, given set $\mathcal{D}_\text{T}$, a demodulator $p_\theta(s|y)$ can be trained by minimizing the cross-entropy loss function:
\begin{align} \label{eq:loss-CE}
L_{\mathcal{D}_\text{T}}(\theta) = -\sum_{(s^{(n)},y^{(n)}) \in \mathcal{D}_\text{T}} \log p_\theta (s^{(n)} | y^{(n)}).
\end{align}
This can be done using Stochastic Gradient Descent (SGD), which iteratively updates the parameter vector $\theta$ based on the rule 
\begin{align} \label{eq:sgd}
\theta \leftarrow \theta -\eta \nabla_\theta \log p_\theta(s^{(n)} | y^{(n)}),
\end{align}
by drawing one pair $(s^{(n)},y^{(n)})$ from the set $\mathcal{D}_{\text{T}}$. In \eqref{eq:sgd}, the step size $\eta$ is assumed to be fixed for simplicity of notation but can in practice be adapted across the updates (see, e.g., \cite{Goodfellow-et-al-2016}). Furthermore, this rule can be generalized by summing the gradient in \eqref{eq:sgd} over a minibatch of pairs from the dataset $\mathcal{D}_{\text{T}}$ at each iteration \cite{Goodfellow-et-al-2016}. 

\begin{figure}
    \centering
    \includegraphics[width=0.85\columnwidth]{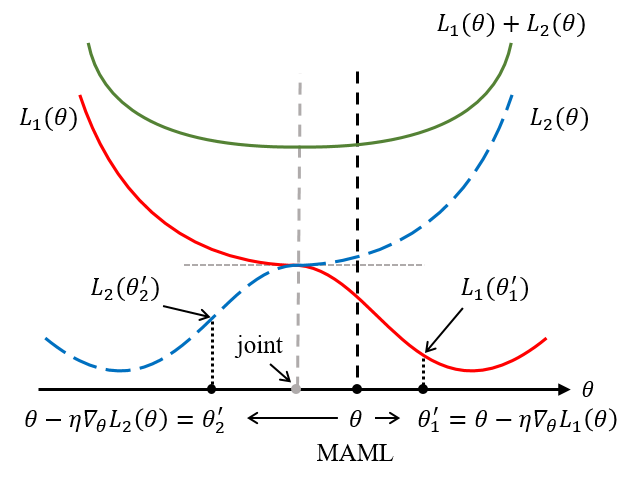}
    \caption{The goal of MAML is to find an initial value $\theta$ that minimizes the loss $L_k(\theta'_k)$ for all devices $k$ after one step of update \eqref{eq:sgd}. In contrast, joint training carries out an optimization on the cumulative loss $L_1(\theta) + L_2(\theta)$.}
    \label{fig:loss_MAML}
\end{figure}

MAML aims at finding an initial parameter $\theta$ such that, starting from this value, the updates \eqref{eq:sgd} are likely to produce a large improvement on the loss function of any target device. Note that it is also possibly advantageous to consider multiple updating steps \eqref{eq:sgd} \cite{antoniou2018train}. This initial parameter is obtained by leveraging the meta-training data $\mathcal{D}$ in the meta-learning phase. As further discussed below, the key idea of MAML is
\newline
\newline

\vspace{-30pt}
\begin{algorithm}[h] 
\DontPrintSemicolon
\LinesNumbered
\smallskip
\KwIn{Meta-training set $\mathcal{D} = \{\mathcal{D}_k\}_{k=1,\ldots,K}$ and pilots $\mathcal{D}_{\text{T}}$ from the target device; $N^\text{tr}$ and $N^\text{te}$; step size hyperparameters $\eta$ and $\kappa$}
\smallskip
\KwOut{Learned parameter vector $\theta$}
\vspace{0.15cm}
\hrule
\vspace{0.15cm}
{\bf initialize} parameter vector $\theta$ \\
\smallskip
\texttt{meta-learning phase} \\
\smallskip
\While{{\em not done}}{
\For{{\em each meta-training device $k$}}{
\smallskip
randomly divide $\mathcal{D}_k$ into two sets $\mathcal{D}_k^{\text{tr}}$ of size $N^\text{tr}$ and $\mathcal{D}_k^{\text{te}}$ of size $N^\text{te}$ \\
\smallskip
evaluate the gradient $\nabla_{\theta} L_{\mathcal{D}_k^{\text{tr}}}(\theta)$ from \eqref{eq:loss-CE} with $\mathcal{D}_\text{T} = \mathcal{D}_k^{\text{tr}}$ \\
\smallskip
compute adapted parameter $\theta_k'$ with gradient-descent \eqref{eq:sgd} of step size $\eta$ as
\begin{align*} 
\theta_k' = \theta - \eta \nabla_\theta L_{\mathcal{D}_k^\text{tr}}(\theta)
\end{align*}
}
\smallskip
update parameter
\begin{align*} 
\theta \leftarrow \theta - \kappa \nabla_\theta \sum_{k=1}^K  L_{\mathcal{D}_k^{\text{te}}}(\theta_k'),
\end{align*}
from \eqref{eq:loss-CE} with $\mathcal{D}_\text{T} = \mathcal{D}_k^{\text{te}}$ for each $k=1,\ldots,K$
}
\smallskip
\texttt{adaptation on the meta-test device} \\
\smallskip
\Repeat{{\em stopping criterion is satisfied}}{
\smallskip
draw a pair $(s^{(n)},y^{(n)})$ from $\mathcal{D}_\text{T}$ \\
\smallskip
update parameter in the direction of the gradient $\nabla_\theta \log p_\theta(s^{(n)} | y^{(n)})$ with step size $\eta$ as in \eqref{eq:sgd} \smallskip
}
\smallskip
\caption{Few-Pilot Demodulator Learning via Meta-Learning}
\label{alg:meta-demodulator}
\end{algorithm}

\noindent to find an initial value $\theta$ such that, for any device, the loss after one iteration \eqref{eq:sgd} applied to the received pilots is minimized. 

To elaborate, assume first that we had available the exact average loss $L_k(\theta)=\mathbb{E}[-\log p_\theta (s|y)]$ for all $k$ meta training devices with $k=1,\ldots,K$. The average in $L_k(\theta)$ is taken over the distribution $p(s,y) = p(s)p(y|s)$, where $p(s)$ is the prior distribution of the transmitted symbol and $p(y|s)$ is defined by \eqref{eq:e2e-channel}. Note that in practice this is not the case since the channel and the non-linearity in \eqref{eq:e2e-channel} are not known a priori. As illustrated in Fig.~\ref{fig:loss_MAML}, MAML seeks a value $\theta$ such that, for every device $k$, the losses $L_k(\theta'_k)$ obtained after one update \eqref{eq:sgd}, i.e., for $\theta'_k = \theta - \eta \nabla_\theta L_k(\theta)$, are collectively minimized. This is done by minimizing the sum $\sum_{k=1}^K L_k(\theta'_k)$ over $\theta$. This procedure can also be interpreted as solving an \emph{empirical Bayes} problem over the prior of the task-specific parameters $\theta'_k$ \cite{grant2018recasting}.

As discussed, the losses $L_k(\theta_k)$ for all meta-training devices are not known and need to be estimated from the available data. To this end, in the meta-learning phase, each set $\mathcal{D}_k$ of $N$ pairs of pilots and received signals for meta-training device $k$ is randomly divided into a training set $\mathcal{D}_k^{\text{tr}}$ of $N^\text{tr}$ pairs and a test set $\mathcal{D}_k^{\text{te}}$ of $N^\text{te}$ pairs, as shown in Fig.~\ref{fig:meta_4pam}. The updated parameter $\theta_k'$ is computed by applying the SGD-based rule in \eqref{eq:sgd} over all pairs in training subset $\mathcal{D}_k^\text{tr}$ as in $\theta_k' = \theta - \eta \nabla_\theta L_{\mathcal{D}_k^\text{tr}}(\theta)$.
The loss $L_k(\theta'_k)$ is then estimated by using the test subset $\mathcal{D}_k^\text{te}$ as $L_{\mathcal{D}_k^\text{te}}(\theta'_k)$. Finally, MAML minimizes the overall estimated loss $\sum_{k=1}^K L_{\mathcal{D}_k^\text{te}}(\theta'_k)$ by performing an SGD-based update in the direction of the gradient $\nabla_\theta \sum_{k=1}^K L_{\mathcal{D}_k^\text{te}}(\theta_k')$ with step size $\kappa$ as 
\begin{align} \label{eq:sgd-meta2}
\theta \leftarrow \theta - \kappa \nabla_\theta \sum_{k=1}^K  L_{\mathcal{D}_k^{\text{te}}}(\theta_k').
\end{align}
The full algorithm is summarized in Algorithm~\ref{alg:meta-demodulator}.

As a benchmark, a more conventional approach to use the meta-training data $\mathcal{D}$ would be to use \emph{joint training} on the meta-training data. Accordingly, one would pool together all the pilots received from the meta-training devices and carry out an SGD-based optimization as in \eqref{eq:sgd} on the cumulative loss $L_{\mathcal{D}}$. In the example in Fig.~\ref{fig:loss_MAML}, this minimization would lead to an initial point that, unlike MAML, would not allow any useful adaptation based on gradient descent. The next section will provide numerical comparisons for the demodulation task under study.

%% file: experiments.tex
\section{Experiments}
\label{sec:exp}

\subsection{Toy Example}
\label{sec:exp-toy}

\begin{figure}
    \centering
    \includegraphics[width=0.9\columnwidth]{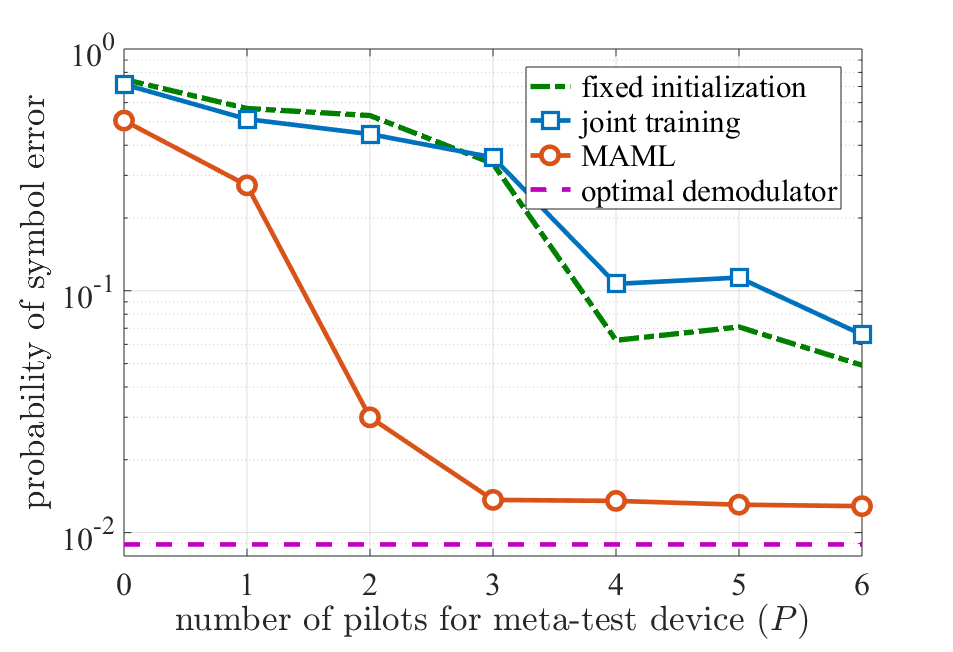}
    \caption{Probability of symbol error with respect to number $P$ of pilots for the meta-test device for an example with binary fading and no amplifier distortion.}
    \label{fig:prob_of_error_simple}
\end{figure}

Before considering a more realistic scenario including channel fading and amplifier's distortion, we start with a simple example, in which we have no amplifier distortion, i.e., $g(s_k) = s_k$ and fading is binary, i.e., the channel $h_k$ in \eqref{eq:e2e-channel} can take values $\pm 1$. This simplified model will be useful to build some intuition about the operation of MAML. We adopt pulse-amplitude modulation with four amplitude levels ($4$-PAM) $\mathcal{S} = \{-3,-1,1,3\}$. Pilot symbols in the meta-training dataset $\mathcal{D}$ and meta-test dataset $\mathcal{D_\text{T}}$ follow a fixed periodic sequence $-3,-1,1,3,-3,-1,\ldots$, while transmitted symbols in the test set for the meta-test device are randomly selected from the set $\mathcal{S}$. The channel of the meta-test device is selected randomly between $+1$ and $-1$ with equal probability, while the channels for half of the meta-training devices are set as $+1$ and for the remaining half as $-1$.

The number of meta-training devices is $K = 20$; the number of pilot symbols per device is $N = 8$, which we divide into $N^\text{tr}=4$ meta-training samples and $N^\text{te}=4$ meta-testing samples. The demodulator \eqref{eq:softmax-demod} is a neural network with $L=3$ layers, i.e., an input layer with $2$ neurons, one hidden layer with $30$ neurons, and a softmax output layer with $4$ neurons, and adopts a hyperbolic tangent function $\sigma(\cdot) = \tanh(\cdot)$ as the activation function. For meta-learning with MAML, we use a minibatch of size $4$ with fixed learning rates $\eta=0.1$ and $\kappa=0.025$. The weights and biases are all initialized to $1$. For the training in meta-test device, we adopt a minibatch of size $1$ and learning rate $\eta = 0.1$. The signal-to-noise ratio (SNR) is given as $\mathbb{E}[s_k^2]/N_o = 15\text{dB}$.

We compare the performance of the proposed MAML approach with: (\emph{i}) a fixed initialization scheme where data from the meta-training devices is not used; (\emph{ii}) joint training with the meta-training dataset $\mathcal{D}$ as explained in Sec.~\ref{sec:meta}; (\emph{iii}) optimal ideal demodulator that assumes perfect channel state information. For (\emph{ii}), we set the learning rate to $0.01$ and the minibatch size to $4$. The probability of error of (\emph{iii}) can be computed as $P_e = \frac{3}{2}Q(\sqrt{\frac{\text{SNR}}{5}})$ using standard arguments. 

\begin{figure}[t]
    \centering
    \includegraphics[width=0.9
    \columnwidth]{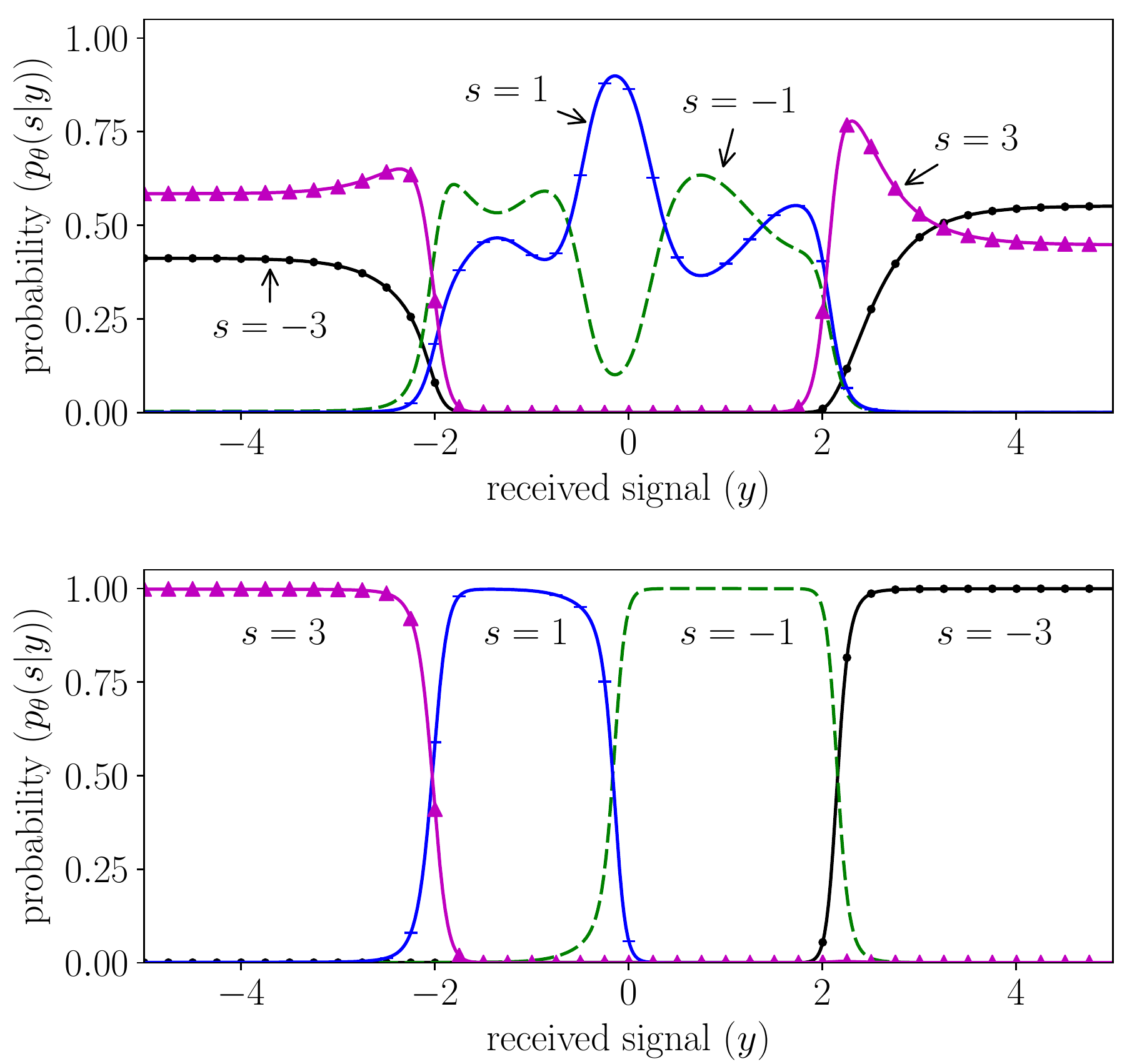}
    \caption{(Top) Demodulator \eqref{eq:softmax-demod} for the parameter vector $\theta$ obtained via meta-learning phase in Algorithm~\ref{alg:meta-demodulator}; (Bottom) Updated demodulator \eqref{eq:softmax-demod} using $P = 6$ pilots from the meta-test device.}
    \label{fig:vis_simple}
\end{figure}

In Fig.~\ref{fig:prob_of_error_simple}, we plot the average probability of symbol error with respect to number $P$ of pilots for the meta-test device. MAML is seen to vastly outperform the mentioned baseline approaches by adapting to the channel of the meta-test device using only a few pilots. In contrast, joint training fails to perform better than fixed initialization. This confirms that, unlike conventional solution, MAML can effectively transfer information from meta-training devices to a new target device. 

In order to gain intuition on how MAML learns from the meta-training devices, in Fig.~\ref{fig:vis_simple}, we plot the probabilities defined by the demodulator \eqref{eq:softmax-demod} for the four symbols in $\mathcal{S}$ with the parameter vector $\theta$ obtained from the meta-learning phase in Algorithm~\ref{alg:meta-demodulator} (top) and after training using the pilots of the target meta-test device (bottom). The class probabilities identified by MAML in the top figure have the interesting property of being approximately symmetric with respect to the origin. This makes the resulting decision region easily adaptable to the channel of the target device, which may take values $\pm 1$ in this example. The adapted probabilities in the bottom figure illustrate how the initial demodulator obtained via MAML is specialized to the channel of the target device.

\subsection{A More Realistic Scenario}
\label{sec:exp-real}

\begin{figure}
    \centering
    \includegraphics[width=0.9\columnwidth]{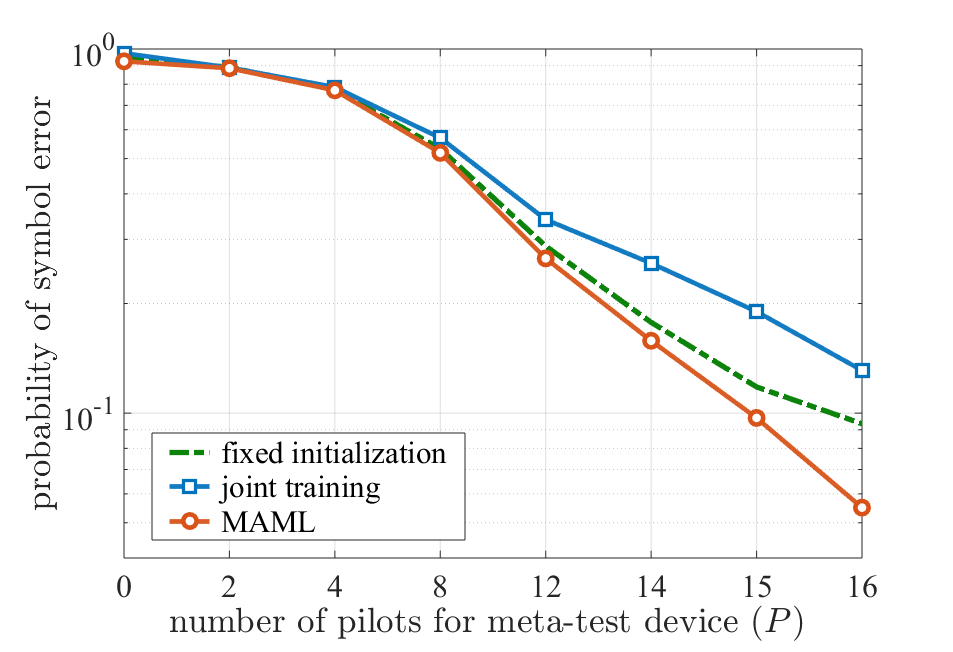}
    \caption{Probability of symbol error with respect to number $P$ of pilots for the meta-test device for $16$-QAM scenario with Rayleigh fading and amplifier distortion. The probability of error of the ideal demodulator with knowledge of channel state information and amplifier's non-linearity transfer function is $P_e = 15Q(\frac{d_{\text{min}}}{\sqrt{2N_o}})$, where $d_{\text{min}} = \min \bigg(\frac{12\sqrt{2}\alpha_k\beta_k A^3-2\sqrt{2}\alpha_k A}{(1+2\beta_k A^2)(1+18\beta_k A^2)}, \frac{2\alpha_k A \sqrt{1+12\beta_k A^2 + 180 \beta_k^2 A^4}}{(1+10\beta_k A^2)(1+18\beta_k A^2)} \bigg)$ with $A$ and $N_o$ determined by $\text{SNR} = \frac{10A^2}{N_o}$.}
    \label{fig:prob_of_error_realistic}
\end{figure}

We now consider a more realistic scenario including Rayleigh fading channels $h_k\sim\mathcal{CN}(0,1)$ and an amplifier's distortion given by $\eqref{eq:non-linearity}$, where $\alpha_k=4$ and $\beta_k$ is uniformly distributed in the interval $[0.05,0.15]$.
We assume $16$-ary quadrature amplitude modulation ($16$-QAM) $\mathcal{S}$ and the sequence of pilot symbols in the meta-training dataset $\mathcal{D}$ and meta-test dataset $\mathcal{D_\text{T}}$ was fixed by cycling through the symbols in $\mathcal{S}$, while the transmitted symbols in the test set for the meta-test device are randomly selected from $\mathcal{S}$. 
The number of meta-training devices is $K=20$; the number of pilot symbols per device is $N=32$, which we divide into $N^{\text{tr}}=16$ meta-training samples and $N^{\text{te}}=16$ meta-testing samples. The demodulator \eqref{eq:softmax-demod} is a neural network with $L=4$ layers, i.e., an input layer with $2$ neurons, two hidden layer with $10$ neurons each, and a softmax output layer with $16$ neurons, and we adopt a rectified linear unit $\sigma(\cdot)=\text{ReLU}(\cdot)$ as the activation function. For meta-learning with MAML, we use a minibatch of size $8$ with fixed learning rates $\eta=0.01$ and $\kappa = 0.001$. The weights and biases are initialized randomly. For the training in meta-test device, we adopt a minibatch of size $8$ if possible or a minibatch of size $1$ if number of pilots is less than $8$, and learning rate $\eta = 0.01$. The signal-to-noise ratio (SNR) is given as $\mathbb{E}[s_k^2]/N_o = 21\text{dB}$.
For the joint training case, we use a minibatch size $8$.

In Fig.~\ref{fig:prob_of_error_realistic}, we plot the average probability of symbol error with respect to number $P$ of pilots in the meta-test device. We compare the performance with fixed initialization and joint training. MAML is seen to adapt more quickly than the baseline schemes to the channel and non-linearity of the target device, although more pilots are needed with respect to the toy example in Fig.~\ref{fig:prob_of_error_simple}.